\documentclass[twocolumn,twocolappendix]{aastex63}
\usepackage{graphicx}
\usepackage{amsmath}

\newcommand{\textred}{\textcolor{black}}
 
\submitjournal{ApJ}

\shorttitle{HSC Weak Lensing Cross Planck CMB Lensing}
\shortauthors{Marques et al. }

\begin{document}

\title{Cross-correlation between Subaru Hyper Suprime-Cam Galaxy Weak Lensing and Planck Cosmic Microwave Background Lensing}

\author[0000-0002-8571-8876]{Gabriela A. Marques}
\affil{Department of Physics, Florida State University, Tallahassee, Florida 32306, USA}

\author[0000-0001-8219-1995]{Jia Liu}
\affiliation{Berkeley Center for Cosmological Physics, University of California, Berkeley, CA 94720, USA}
\affiliation{Lawrence Berkeley National Laboratory, 1 Cyclotron Road, Berkeley, CA 93720, USA}
 
\author[0000-0001-7109-0099]{Kevin M.\ Huffenberger}
\affiliation{Department of Physics, Florida State University, Tallahassee, Florida 32306, USA}
 
\author[0000-0002-9539-0835]{J.~Colin Hill}
\affiliation{Department of Physics, Columbia University, New York, NY, USA 10027}
\affiliation{Center for Computational Astrophysics, Flatiron Institute, New York, NY, USA 10010}
 
\correspondingauthor{Gabriela A. Marques}
\email{gmarques@fsu.edu}

\begin{abstract}
Cross-correlations between galaxy weak lensing (WL) and Cosmic Microwave Background (CMB) lensing are powerful tools to probe matter fluctuations at intermediate redshifts and to detect residual systematics in either probe. In this paper, we study the cross-correlation of galaxy WL from the Hyper Suprime-Cam Subaru Strategic Program (HSC) first data release and  CMB lensing from the final Planck data release, for HSC source galaxies at $0.3\leq z\leq1.5$. 
HSC is the deepest Stage-III galaxy WL survey, and provides a great opportunity to study the high-redshift universe. It also presents new challenges related to its exceptionally high source density, such as source blending.
The cross-correlation signal is measured at a significance level of $3.1\sigma$. The amplitude of our best-fit model with respect to the best-fit 2018 Planck cosmology is $A = 0.81\pm 0.25$, consistent with $A=1$. Our result is also consistent with previous CMB lensing and galaxy WL cross-correlation studies using different surveys. We perform tests with respect to the WL $B$-modes, the point-spread-function, photometric redshift errors, and thermal  Sunyaev-Zel'dovich leakage, and find no significant evidence of residual systematics.

\end{abstract}

\section{Introduction}
\label{sec:intro}
The gravitational potential of large-scale structure deflects the path of photons as they traverse the Universe --- an effect known as weak gravitational lensing~(WL). Weak lensing of the Cosmic Microwave Background (hereafter, CMB lensing) has matured from early detection to a standard cosmological probe in the past 15 years~\citep{hirata2004cross,smith2007detection,sherwin2012atacama, das2014atacama,ade2014measurement, lensing2018planck}. Using photons from the last-scattering surface as a back light, CMB lensing measures the matter distribution early in the history of structure growth. Weak lensing of galaxies (hereafter, galaxy WL), using galaxies instead as the source, has also achieved competitive constraints on cosmology in the recent years by multiple experiments~\citep{heymans2012cfhtlens,de2013kilo,erben2013cfhtlens,shan2014weak, becker2016cosmic,hildebrandt2016rcslens,jee2016cosmic,zuntz2018dark}. With the galaxies distributed at a wide range of redshifts, galaxy WL is a powerful tool to build a tomographic model of the growth of structure. 

We expect the signals from CMB lensing and galaxy WL to be correlated when the same patch of sky is observed, as their lensing kernels overlap in redshift. The cross-correlation of CMB lensing and galaxy WL is sensitive to the underlying matter distribution and therefore can be used to constrain cosmological parameters, such as the mass fluctuation amplitude $\sigma_{8}$ and the matter density $\Omega_{\rm m}$. In addition, because CMB and galaxy surveys have uncorrelated noise, the correlation also provides a consistency check of systematic residuals in individual surveys, such as the shear multiplicative bias and photometric redshift (photo-$z$) errors in galaxy WL surveys, and foreground contamination from dust, the cosmic infrared background, or the Sunyaev-Zel'dovich effects in CMB lensing maps.
 
The CMB lensing and galaxy WL cross-correlation has been studied with many surveys over the past several years, using CMB lensing data from the Planck mission, the Atacama Cosmology Telescope (ACT), and the South Pole Telescope (SPT) and galaxy WL data from the CFHT Stripe 82 Survey (CS82), Canada-France-Hawaii Telescope Lensing Survey (CFHTLenS),  Sloan Digital Sky Survey (SDSS), Red Cluster Sequence Lensing Survey (RCSLenS), Kilo Degree Survey (KiDS), and Dark Energy Survey (DES) ~\citep{hand2015first,liu2015cross,harnois2016cfhtlens,singh2017cross,harnois2017kids,omori2019dark,kirk2016cross}. To date, CMB lensing maps used in such analyses are typically reconstructed using CMB temperature anisotropy data, or that jointly with CMB polarization data. Recently, the first detection of such cross-correlation using CMB lensing maps reconstructed exclusively from CMB polarization data has been reported by the Polarbear experiment and the HSC Collaboration~\citep{namikawa2019evidence}. 
 
In this work, we study the cross-correlation between Planck 2018 CMB lensing data~\citep{lensing2018planck} and the six galaxy WL fields from the HSC first year shear catalog~\citep{mandelbaum2018weakhsc, oguri2018two}. As the deepest Stage-III galaxy WL  survey, HSC can be considered a path-finder for the upcoming Vera Rubin Observatory ~\citep{hikage2019cosmology,nicola2020tomographic,osato2020cross,makiya2020new,chiu2019richness}. As HSC can observe galaxies at higher redshift, this cross-correlation is sensitive to structure growth earlier in time ($z\approx1$) than probed in previous works. The high number density achieved by HSC also presents new challenges to galaxy WL analysis, such as source blending and masking. In addition, cross-correlating the HSC galaxy WL data with CMB lensing provides a possibility to probe residual systematics and biases present in galaxy weak lensing, effects that are difficult to model through auto-correlation analysis~\citep{vallinotto2012using}. It is also important to check our results against previous works, in particular works using Planck data.

This paper is organized as follows. In Sec.~\ref{sec:th}, we review the theoretical background. We then describe the data used in our analysis in Sec.~\ref{sec:data} and the method to measure the cross-correlation in Sec.~\ref{sec:method}. We present our results in Sec.~\ref{sec:results} and null tests in Sec.~\ref{sec:null_tests}. Finally, we conclude in Sec.~\ref{sec:conclusions}.
   
\section{Formalism} 
\label{sec:th}
Weak lensing signals are detected as distortions in galaxy shapes (``shear'') in galaxy WL~\citep{bartelmann2001weak,refregier2003weak,mandelbaum2018weak} or an \textred{isotropy violation} in the CMB temperature and polarization maps in CMB lensing~\citep{lewis2006weak,hanson2010weak}. Under the Born approximation, where photons travel along the unperturbed path, and assuming a flat universe, the lensing potential is the line-of-sight integral over the Weyl potential $\Psi$,
\begin{align}
    \Phi(\boldsymbol{\theta},\chi)  = \frac{2}{c^2}\int_{0}^{\chi}\frac{\chi_{\rm s} - \chi_{\rm l}}{\chi_{\rm s}\chi_{\rm l}} \Psi(\chi_{\rm l}\boldsymbol{\theta}, \chi_{\rm l}),
    \label{eq:lens_potential}
\end{align}
where $\boldsymbol{\theta}$ is the angular coordinate, $c$ is the speed of light, and $\chi_{\rm s}$ and $\chi_{\rm l}$ are the comoving radial  distances to the source and the lens, respectively. 

The lensing signal is typically quantified as a convergence field $\kappa$ for CMB lensing and as a complex shear field $\gamma = \gamma_{1}+i\gamma_{2}$ for galaxy WL. However, $\kappa$ and $\gamma$ are not independent. They are both related to the second derivatives of the lensing potential,
\begin{align}
\kappa = & \frac{1}{2} \nabla^2\Phi ; \\
\gamma_{1} =& \frac{1}{2}(\partial_{1}\partial_{1} - \partial_{2}\partial_{2})\Phi ; \\
\gamma_{2} =& \partial_{1}\partial_{2}\Phi,
\end{align}
where the partial derivatives $\partial_{1},\partial_{2}$ are with respect to the plane-of-sky coordinates. The shear field can be decomposed into its divergence (E-mode) and curl (B-mode) components, using e.g.\ the Kaiser~\&~Squires inversion method~\citep{kaiser1993mapping}. Since gravitational potentials can only generate E-mode signals, a non-zero B-mode signal can point to systematic effects in the data~\citep{krause2010weak}.\ 

To quantify the correlation between CMB lensing and galaxy WL, we use the angular cross-power spectrum between the CMB convergence map $\kappa_{\rm{cmb}}$ and the E-mode shear field $\gamma_{E}$. Under the Limber approximation \citep{limber1953analysis},
\begin{align}    
    C_{\ell}^{\kappa_{\rm{cmb}}\gamma_{\rm{E}}} =
    \int_{0}^{\chi_{*}}d\chi\frac{W^{\kappa_{\rm{cmb}}}(\chi)W^{\gamma_{\rm{E}}}(\chi)}{\chi^2}P\left(k,z \right),
    \label{eq:limber_cross}
\end{align}
where $P(k,z)$ is the non-linear matter power spectrum at wavenumber $k$=$(\ell+\frac{1}{2})/{\chi}$ and redshift $z$=$z(\chi)$, $\chi_{*}$ is the comoving distance to the last scattering surface, and $W^{\gamma_{\rm{E}}}$ and $W^{\kappa_{\rm{cmb}}}$ are the galaxy WL and the CMB lensing window functions, respectively,  
 \begin{align}
     W^{\gamma_{\rm{E}}}(\chi) =& \frac{3\Omega_{\rm m}H_{0}^2}{2c^2}\frac{\chi}{a(\chi)}\int_{\chi}^{\infty}d\chi'p_{\rm z}(\chi')\frac{\chi'-\chi}{\chi'}, \label{eq:kernel_gal}
\\
     W^{\kappa_{\rm{cmb}}}(\chi) =& \frac{3\Omega_{\rm m}H_{0}^2}{2c^2}\frac{\chi}{a(\chi)}\frac{\chi_{*}-\chi}{\chi_{*}},
     \label{eq:kernel_cmb}
 \end{align}
where $\Omega_{\rm m}$ is the present-day matter density, $H_{0}$ is the Hubble constant, $a(\chi)$ is the scale factor, and $p_{\rm z}(\chi')$ is the redshift distribution of source galaxies normalized such that $\int d\chi'p_{\rm z}(\chi') =1$.

If we assume both fields are Gaussian, the theoretical variance on the cross-power spectrum is,
\begin{align}
    (\sigma_{\ell}^{\kappa_{\rm{cmb}}\gamma_{\rm{E}}})^2 = & \frac{1}{(2\ell+1)f_{\rm sky}\Delta\ell} \nonumber\\  
    \times & \left[(C_{\ell}^{\kappa_{\rm{cmb}}\gamma_{\rm{E}}})^2+ C_{\ell}^{\gamma_{\rm{E}}\gamma_{\rm{E}}}C_{\ell}^{\kappa_{\rm{cmb}}\kappa_{\rm{cmb}}}\right],
    \label{eq:errorcl}
\end{align}
where $\Delta \ell$ is the bin width, $f_{\rm sky}$ is the sky fraction of the overlapping area of the surveys,  $C_{\ell}^{\gamma_{\rm{E}}\gamma_{\rm{E}}}$  and $C_{\ell}^{\kappa_{\rm{cmb}}\kappa_{\rm{cmb}}}$ are the auto-spectra of the galaxy WL and CMB lensing, respectively. In our analysis, we use simulated maps to estimate the full covariance and only use Eq.~\ref{eq:errorcl} to validate our results.

We calculate $P(k,z)$ using the the \texttt{Halofit} model \citep{smith2003stable,takahashi2012} implemented in the Boltzmann code \texttt{CAMB}\footnote{\url{https://camb.info/}}\citep{lewis2011camb}. Throughout the paper, we assume the best-fit Planck 2018 cosmology (\texttt{TT,TE,EE+lowE+CMB lensing})~\citep{planck2018params}: $\{\Omega_{\rm b}h^2, \Omega_{\rm c}h^2, \Omega_{\rm m}, \tau, n_{\rm s}, A_{\rm s},h\}$=$\{0.0223,  0.1200,$ $0.3153,0.054, 0.964, 2.1\times 10^{-9}, 0.673\}$.
 
\section{Data}
\label{sec:data}

\subsection{HSC Galaxy Weak Lensing}
\label{sec:data_hsc}
We use the HSC first-year shear catalog~\citep{mandelbaum2018first} (hereafter, S16A), covering a 136.9~$\deg^{2}$ sky region with a limiting magnitude $i_{\rm lim}$=24.5. The galaxy shapes are measured from high-quality $i$-band images with a median point-spread function~(PSF) FWHM~$\approx0.58$ arcsec \citep{bosch2018hyper}.
The total shear catalog contains more than 12 million galaxies in 6 disjoint regions: \texttt{XMM}, \texttt{GAMA09H}, \texttt{WIDE12H}, \texttt{GAMA15H}, \texttt{VVDS}, and  \texttt{HECTOMAP}. 
The shape $\boldsymbol{e}=(e_{1},e_{2})$ of the galaxies are estimated on the coadded $i$-band images using the re-Gaussianization PSF correction method~\citep{hirata2003shear}. In addition, the catalog provides the additive biases $\boldsymbol{c}=(c_{1},c_{2})$,  multiplicative bias $m$, the intrinsic shape root mean square per component $e_{\rm rms}$, and the weight $w$, defined as the inverse variance of the shape noise
\begin{equation}
    w = (\sigma^2_{\rm e} + e^2_{\rm rms})^{-1},
\end{equation}
where $\sigma^2_{\rm e}$ represents the shape measurement error for each galaxy.
The smoothed shear map $\hat{\gamma}_{\alpha}$ ($\alpha=1,2)$ can be constructed using
\begin{align}
\hat{\gamma}_{\alpha}({\boldsymbol\theta}) = & \frac{\sum_{i}w_{i}[\gamma_{\alpha}({\boldsymbol\theta_{i}})-c_{\alpha,i}]W_{\rm G}(\mid\boldsymbol{\theta}-\boldsymbol{\theta}_{i}\mid)}{\sum_{i}w_{i}(1+m_{i})W_{\rm G}(\mid \boldsymbol{\theta} - \boldsymbol{\theta}_{i} \mid)}
\end{align}
where $i$ runs over all galaxies, and $W_{\rm G}$ is a Gaussian smoothing kernel
\begin{equation}
    W_{G}(\theta) = \frac{1}{\pi\theta_{s}^2}\exp{\bigg(-\frac{\theta^2}{\theta_{s}^2} \bigg)},
\end{equation}
with a smoothing scale $\theta_{s}=$1 arcmin. The shear $\gamma_{\alpha}$ is related to the shape measurement $e_{\alpha}$ through a shear responsivity $\mathcal{R}$ ,
\begin{align}
\gamma_{\alpha}({\boldsymbol\theta_{i}}) =& \frac{e_{\alpha}({\boldsymbol\theta_{i}})}{2\mathcal{R}},\\
\mathcal{R} = & 1-\frac{\sum_{i}w_{i}e^2_{{\rm rms},i}}{\sum_{i}w_{i}}.   
\end{align}

Given the small size of each HSC field, we adopt a flat-sky approximation. Our pixelated map has a 0.88 arcmin resolution, following \citet{hikage2019cosmology}. The mask for each field is constructed by setting pixels with non-zero weights to 1 and 0 otherwise, where the weight in each pixel is the sum of weights of galaxies within that pixel. In addition, following \citet{oguri2018two}, we construct a smoothed number density map using the same $W_{\rm G}$ kernel and remove pixels with a number density smaller than half of the mean number density. This removes regions corresponding to edges, low density pixels and regions that are affected by bright objects such as stars.

\begin{figure*}
	\centering
	\includegraphics[width=\linewidth]{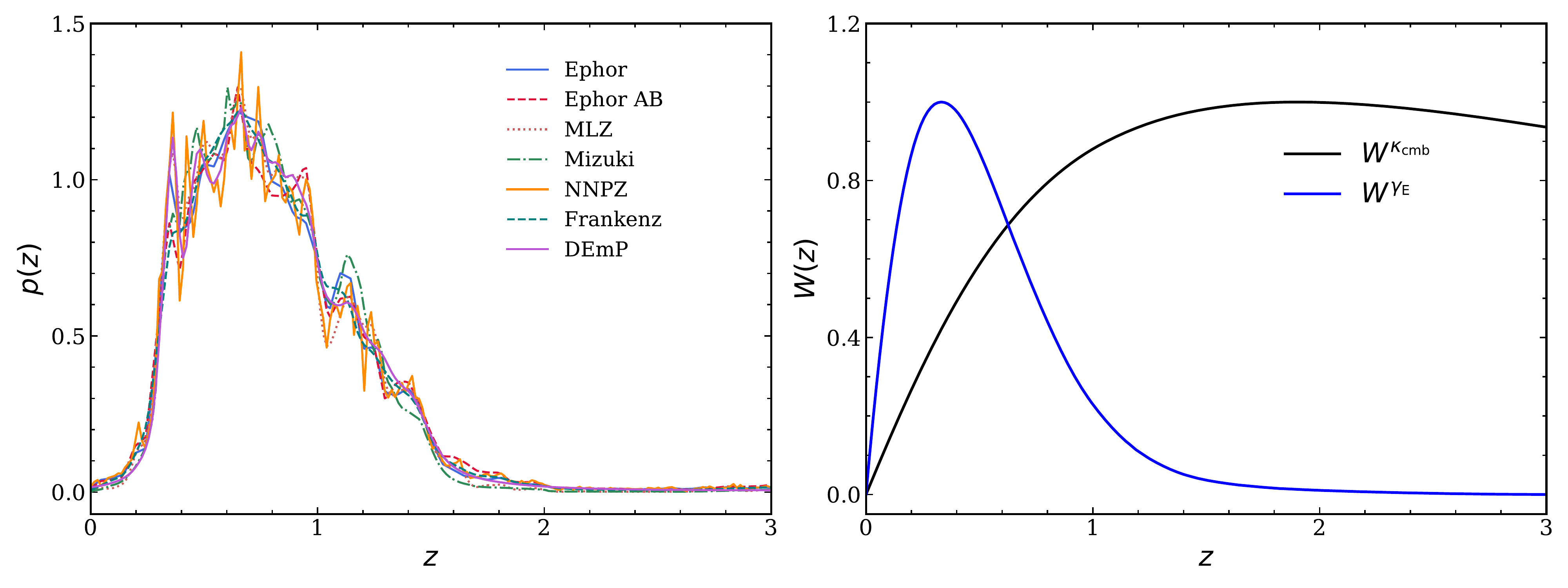}
	\caption{ {\bf Left panel}: HSC galaxy redshift distribution obtained from stacking the photo-$z$ PDFs using different photo-$z$ codes: \texttt{MLZ, Ephor, Ephor AB, Mizuki, NNPZ, Frankenz} and \texttt{DEmP}~\citep{tanaka2018photometric}. We adopt the \texttt{Ephor AB} in our baseline analysis. We apply a redshift cut $0.3<z_{\rm best}<1.5$, where $z_{\rm best}$ is the best-fit photometric redshift. {\bf Right panel}: The galaxy weak lensing kernel from \texttt{Ephor AB} (blue line, Eq.~\ref{eq:kernel_gal}) and the CMB lensing kernel (black line, Eq.~\ref{eq:kernel_cmb}). }
\label{fig:pz_stacked}
\end{figure*}

To estimate the theoretical cross-power spectrum, we need the redshift distribution of the source galaxies. The photometric redshifts of the HSC galaxies are measured using several photo-$z$ algorithms, namely \texttt{MLZ, Ephor, Ephor AB, Mizuki, NNPZ, Frankenz} and \texttt{DEmP} (see \cite{tanaka2018photometric} for details). 
Fig.~\ref{fig:pz_stacked} shows the stacked photo-$z$  probability distribution function (PDF) from each algorithm.
Throughout our analysis, we use the stacked photo-$z$ PDF estimated from \texttt{Ephor AB}, following the HSC cosmic shear auto-power spectra analysis~\citep{hikage2019cosmology}. We test our pipeline using other photo-$z$ algorithms and verify in Sec.~\ref{sec:null_tests} that this choice does not impact significantly our results. 

We restrict the source redshift range to $0.3<z_{\rm best}<1.5$, where $z_{\rm best}$ is the best-fit photo-$z$ determined by \texttt{Ephor AB} (see section 4.2 of \cite{tanaka2018photometric}). The final shape catalog contains $\sim 9$ million galaxies after this cut in redshift, with a mean redshift of $\sim 0.8$. 
The stacked photo-$z$ PDFs shown in Fig.~\ref{fig:pz_stacked} show tails that extend beyond our redshift cut, because of the photo-z uncertainties. We also show the galaxy WL kernel (only shown for \texttt{Ephor AB} for clarity) and the CMB lensing kernel in Fig.~\ref{fig:pz_stacked}. The galaxy WL kernel peaks at around redshift of 0.4 for the current data selection, while the CMB lensing kernel peaks at a redshift $\sim 2$. 
The cross-correlation allows us to probe the large-scale structure at an intermediate redshift since the joint kernel peaks at $\langle z\rangle = \int_{0}^{\infty} W^{\rm \kappa_{cmb}} W^{\gamma_{E}} zdz/ \int_{0}^{\infty}W^{\rm \kappa_{cmb}} W^{\gamma_{E}} \approx 0.8$.\

\begin{table}
\begin{tabular}{lcrc}
\hline
HSC Field    & Area  & $N_{\rm total}$ & $n_{\rm gal}^{\rm eff}$ \\ 
        & ($\deg^2$)&&(arcmin$^{-2}$)\\
\hline
\texttt{WIDE12H}  & 12.82          & 714,058            & 19.96                         \\
\texttt{HECTOMAP} & 10.68           & 565,615             & 19.12                          \\
\texttt{GAMA09H}  & 35.00           & 1,640,415             & 17.23                          \\
\texttt{GAMA15H}  & 25.47           & 1,384,267            & 23.24                          \\
\texttt{VVDS}     & 19.24           & 1,022,618        & 19.42                          \\
\texttt{XMM}      & 28.60           &1,438,848            & 18.06                          \\ \hline
\end{tabular}
\caption{The total area, total number of source galaxies, and the effective number density of each HSC field. The maps are smoothed with a 1 arcmin Gaussian window. Joint HSC and Planck masks are applied.}\label{tab:fields_caract}

\end{table}

We summarize in Table~\ref{tab:fields_caract} the area, number of galaxies, and the effective number density for each smoothed, masked HSC field. The effective number density as defined by \citet{heymans2012cfhtlens} is
\begin{align}
n_{\rm gal}^{\rm eff} = \frac{1}{\Omega_{\rm sky}}\frac{(\sum_{i}{w_{i}})^2}{\sum_{i}w_{i}^2},
\end{align}
where $\Omega_{\rm sky}$ is the sky area. We use a joint HSC and Planck mask (see below).

\subsection{Planck CMB Lensing}
We use the public 2018 Planck lensing potential  maps\footnote{\url{PLA:https://pla.esac.esa.int/}}~\citep{lensing2018planck} reconstructed using quadratic estimators~\citep{okamoto2003cosmic}. In our main analysis, we use the \texttt{COM\_Lensing\_4096\_R3.00} map, obtained from a minimum-variance (MV) combination of the multi-frequency, foreground-reduced \texttt{SMICA} temperature and polarization maps~\citep{Akrami:2018mcd}. The lensing maps are released in harmonic space, with coefficients up to $\ell_{\rm max} = 4096$. However, because the map is noise-dominated at the higher multipoles  (smaller scales), we limit our analysis to $\ell_{\rm max} = 2048$. The map is generated using the HEALPix pixelisation scheme \citep{gorski2005healpix}, with resolution parameter N$_{\rm side}$ = 4096.

We apply the associated lensing mask in order to attenuate the foreground contamination, leaving a total unmasked sky fraction of $f_{\rm sky}= 0.671$. 
However, small systematic biases due to residual foregrounds may still affect the lensing maps even after applying this mask (and the \texttt{SMICA} cleaning), such as those due to the kinetic Sunyaev-Zel'dovich (kSZ) effect, thermal Sunyaev-Zel'dovich (tSZ) effect, and Cosmic Infrared Background (CIB)~\citep{van2014cmb,Osborne:2013nna,hojjati2017cross,madhavacheril2018mitigating,baxter2019dark,schaan2019foreground}. 
At the sensitivity level of Planck, the kSZ contamination is expected to be subdominant and therefore statistically negligible in our analysis~\citep{ferraro2018bias}. However, the tSZ signal could be directly correlated with the lensing potential and with the large scale structure tracers. To assess the possible impact of the tSZ effect, we repeat our analysis using the \texttt{COM\_Lensing-Szdeproj\_4096\_R3.00} map, reconstructed from the tSZ-deprojected \texttt{SMICA} temperature map, and confirm that our main results are unaffected (Sec.~\ref{sec:null_tests}). At the same time, this test gives us further confidence that there is no significant CIB contamination since the expected CIB-induced biases are considerably different in the tSZ-deprojected \texttt{SMICA} weighting, as shown in Fig. 23 in \cite{lensing2018planck}.
\begin{figure*}[!htbp]
	\centering
	\includegraphics[width=.7\linewidth]{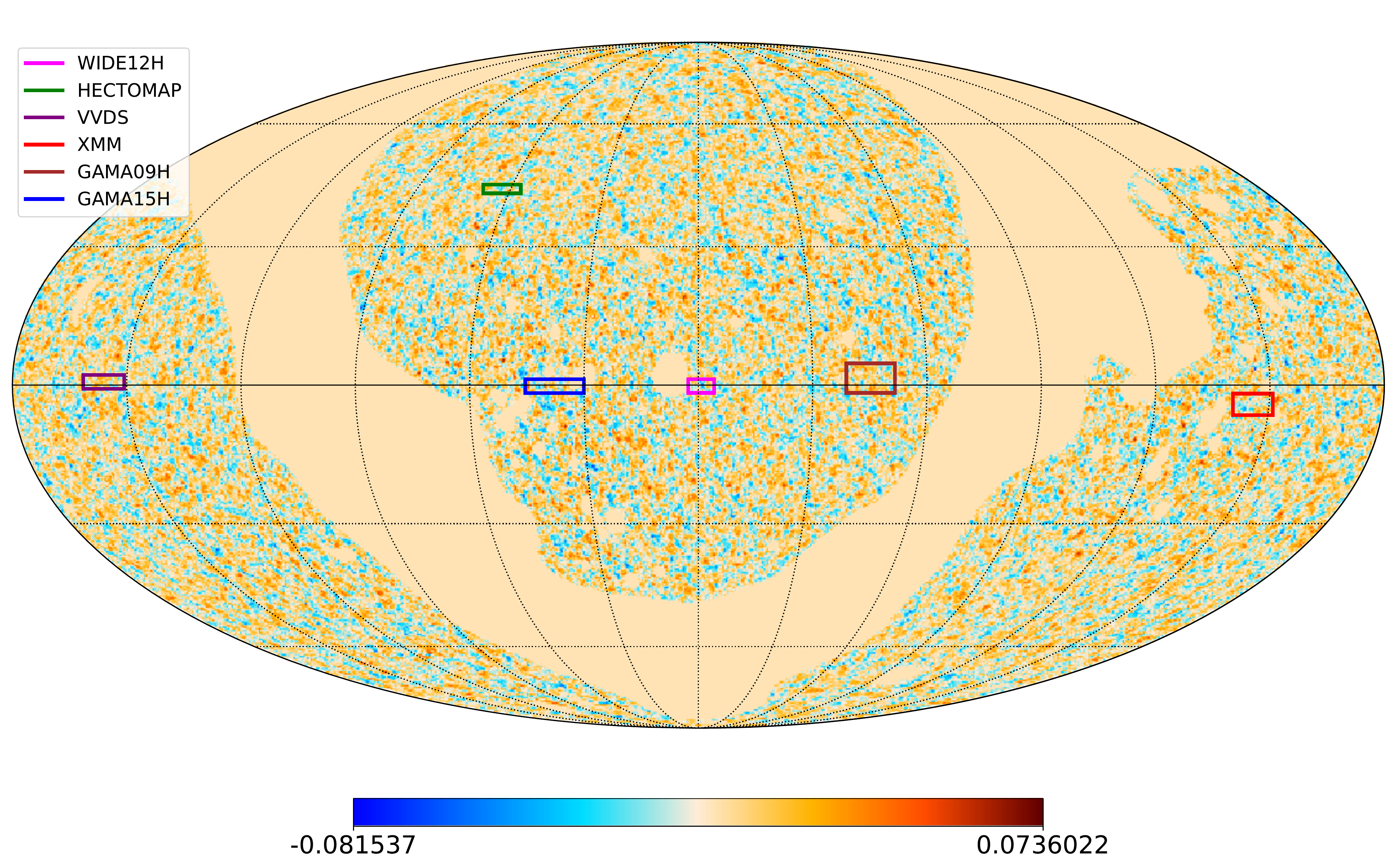}
	\caption{The reconstructed Planck CMB lensing map overlaid with the footprints of the six HSC fields: \texttt{XMM, WIDE12H, Hectomap, VVDS, GAMA15H}, and \texttt{GAMA09H}. We use the Mollweide projection and equatorial coordinates centering at R.A.=12h. For visual purposes, we applied a Wiener filter to the CMB lensing map.}
\label{fig:hsc_fields}
\end{figure*}

We select the HSC regions on the HEALPix CMB lensing map and project it onto the cylindrical equal-area coordinates using the \texttt{flipper} software\footnote{\url{https://github.com/sudeepdas/flipper}}~\citep{das2009efficient}. We then combine the Planck and HSC masks and we apodize the total mask, following the procedure for smooth apodization using \texttt{NaMaster}\footnote{\url{https://github.com/LSSTDESC/NaMaster}} software \citep{alonso2019unified}. This method set to zero all pixels inside a radius of 2.5 times the apodization scale, $\theta_{\rm m}$, of a masked pixel. Then, the mask is smoothed with a Gaussian kernel with a standard deviation equals to $\theta_{\rm m}$ and the pixels originally masked are set to zero again. Consequently, the original masked regions will remain the same, while their edges will have a smooth transition from zero to one. We considered an apodization scale of $\theta_{\rm m} = 0.3'$ and apply the obtained mask to all datasets. Fig.~\ref{fig:hsc_fields} shows the reconstructed Planck CMB lensing map and the footprints of the six HSC fields.

\section{Methods}
\label{sec:method}
\subsection{Power Spectrum estimation}
We estimate the cross-power spectra using the pseudo-$C_{\ell}$ approach with the \texttt{NaMaster} code ~\citep{alonso2019unified}. 
We first measure the cross-power spectra directly from observation, 
\begin{align}
C_{\ell}^{{\rm obs,}\kappa_{\rm{cmb}}\gamma_{\rm{E}}}=
\langle \kappa_{\rm{cmb}}({\boldsymbol\ell})\gamma_{\rm{E}} ^*({\boldsymbol\ell}) \rangle_\ell
\end{align} 
where $\gamma_{\rm E}({\boldsymbol\ell})$ and 
$\kappa_{\rm cmb}({\boldsymbol\ell})$ are the observed and masked galaxy WL and CMB lensing maps in Fourier space, respectively, and $\ell=|\boldsymbol\ell|$.

The observed cross-power spectrum is related to the underlying true power spectrum ($C_{\ell}^{\kappa_{\rm{cmb}}\gamma_{\rm{E}}}$) through 
\begin{align}
   C_{\ell}^{{\rm obs,}\kappa_{\rm{cmb}}\gamma_{\rm{E}}} = \sum_{\ell'}M_{\ell \ell'}C_{\ell'}^{\kappa_{\rm{cmb}}\gamma_{\rm{E}}}. 
\end{align}
The matrix $M_{\ell \ell'}$ describes the mode-coupling introduced by the survey masks and can be computed analytically~\citep{hivon2001master}. We invert $M_{\ell \ell'}$ to obtain the true cross-power spectra.

We compute the cross-power spectrum for each of the six HSC fields, between the multipole range $2 <\ell< 2400$. However, due to the limited area covered by the HSC and the noise in the Planck CMB lensing map, we limit our analysis for  5 linearly spaced bins between $100 <\ell< 1900$.
Although we did not use the lower ($\ell <100$) and the higher multipoles ($\ell > 1900$) in the analysis, we include these bins to perform the inversion of the binned coupling matrix to prevent the bias from these scales. 
 
\subsection{Covariance Matrix}
\label{ref:covariance}
The CMB lensing map from Planck is noise dominated on most scales~\citep{lensing2018planck}. In comparison with the other components in Eq.~\ref{eq:errorcl}, the CMB lensing auto-power spectrum and its noise should dominate the variance of the cross-power spectrum. We evaluate the covariance matrix from the cross-correlation measurements between the real HSC fields and a set that accurately reflects the CMB lensing signal and noise properties, also available from Planck collaboration~\citep{lensing2018planck}. We measure the cross-power spectra between 300 simulated Planck lensing maps and the real HSC galaxy WL fields, carrying out the same approach as the performed for the data. 

This choice ignores the cross-term in Eq.~\ref{eq:errorcl}, which we expect to be negligible compared to the noise contribution from the Planck lensing auto-power spectrum. To test this assumption we calculate the Gaussian prediction by inserting the observed power spectrum of each component in Eq.~\ref{eq:errorcl}, so that these predictions naturally take into account the statistical noise and the effect of masks. We found a very good agreement between the diagonal components of our covariance matrix and the theoretical variance, within 10\% over the angular scales of interest for all six fields. We also estimate the covariance using a data-based jackknife method, where data is resampled from 50 equal-area regions. The diagonal components estimated from this method are on average $\sim 15\%$ larger than those of our simulation-based covariance. Given that the jackknife method is known to overestimate errors \citep{norberg2009statistical,friedrich2016performance}, this result provides a reference for the upper bound of the true covariance and is consistent with our simulation-based method. 

The covariance matrix is dominated by its diagonal components. The values of off-diagonal components are $\lesssim10\%$ of the diagonal components. Nevertheless, we use the full covariance in our analysis. 

\section{Results}
\label{sec:results}

\begin{figure}
	\centering
	\includegraphics[width=\linewidth]{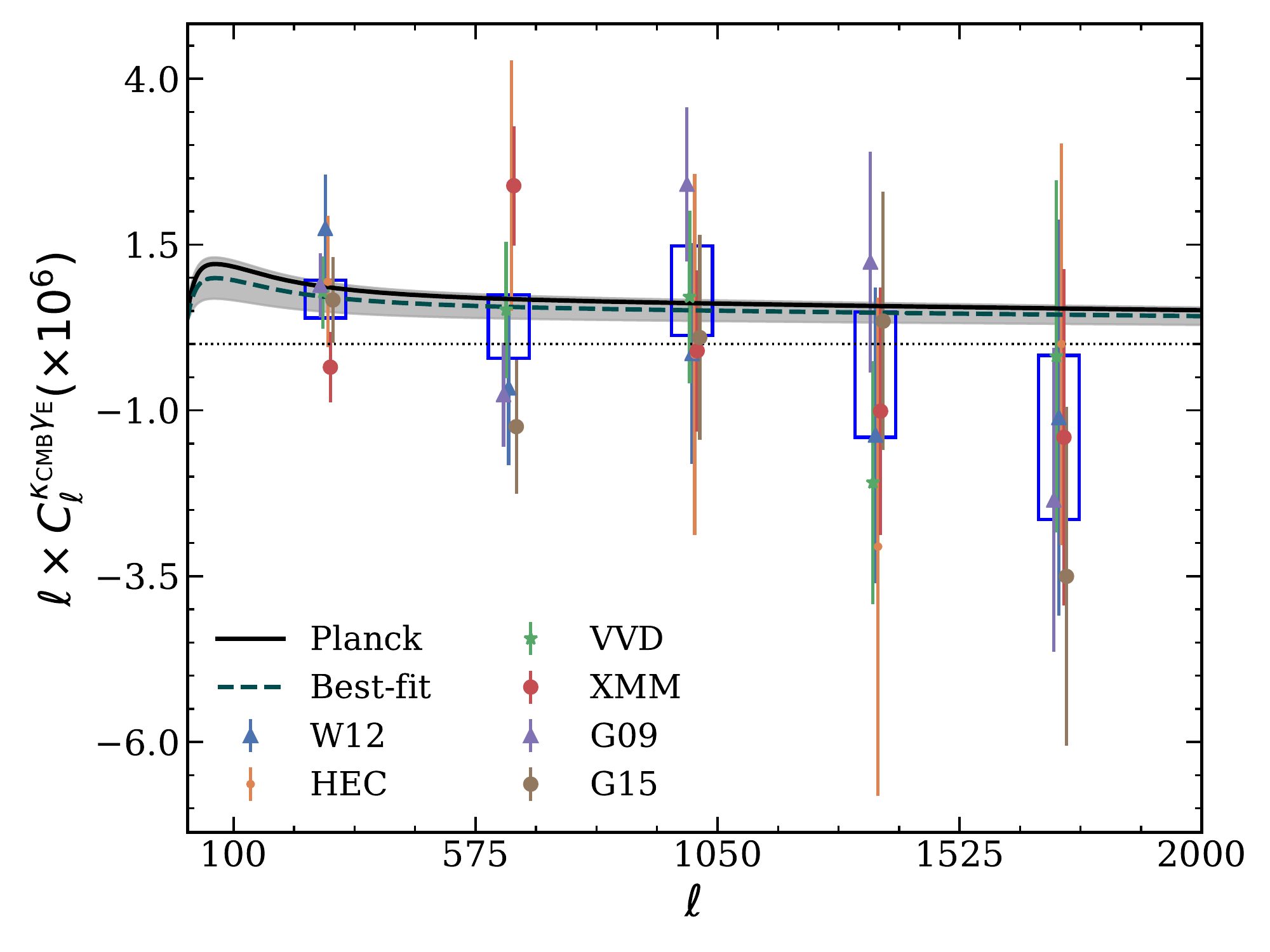}
	\caption{Cross-correlation between the Planck CMB lensing map and the HSC galaxy weak lensing maps. The data points are for individual fields. The theoretical prediction (black solid line) assumes Planck 2018 cosmology (i.e, $A=1$). The best-fit amplitude with respect to Planck 2018 cosmology is $A= 0.81\pm 0.25$ (dashed line) with $1\sigma$ errors shown in shaded gray region. The boxes represent the inverse-variance weighted sum of the measurements of the six fields.}
\label{fig:cls_cross_result}
\end{figure}

We show the angular cross-power spectra between Planck CMB lensing and the six HSC galaxy WL maps in Figure~\ref{fig:cls_cross_result}. The theoretical prediction assuming Planck 2018 cosmology is also shown~(solid black line). The error bars are obtained from the diagonal components of the covariance matrix~(Sec.~\ref{ref:covariance}). 
We fit a best model to our data using an amplitude parameter $A$ by minimizing:
\begin{align}
    \chi^2 = (\mathbf{d} -A \mathbf{t})^{T}\mathrm{Cov}^{-1}(\mathbf{d} -A \mathbf{t})
\end{align}
where $\mathbf{d}$ is the data vector, $\mathbf{t}$ is theoretical prediction assuming Planck 2018 cosmology with the same binning applied, and $\mathrm{Cov}^{-1}$ is the inverse covariance matrix. We apply a pre-factor  $(N_{\rm  sim}-N_{\rm bin}-2)/(N_{\rm sim}-1)$ to $\mathrm{Cov}^{-1}$ to correct for biases introduced by the limited number of simulations~\citep{hartlap2007your}, where $N_{\rm sim}$=300 is the number of simulations used and $N_{\rm bin}$=30 is the number of bins (5 bins $\times$ 6 HSC fields). If the data match the model, then $A=1$.

Our best-fit amplitude is $A= 0.81\pm 0.25$, with $\chi_{\rm min}^2 = 31.2$ for 29 degrees of freedom (DOF), corresponding to a probability-to-exceed PTE=35.3$\%$. We found the significance of the detection SNR$=\sqrt{\chi^2_{\rm null}-\chi_{\rm min}^2}$= 3.1, where $\chi^2_{\rm null}$ is computed by setting $A=0$. The best-fit model is also shown in Fig.~\ref{fig:cls_cross_result} (dashed line) with the corresponding $1\sigma$ errors (gray shaded region). The blue boxes represent the inverse-variance weighted sum of the six fields.
                      
\begin{figure*}
	\centering
	\includegraphics[width=\linewidth]{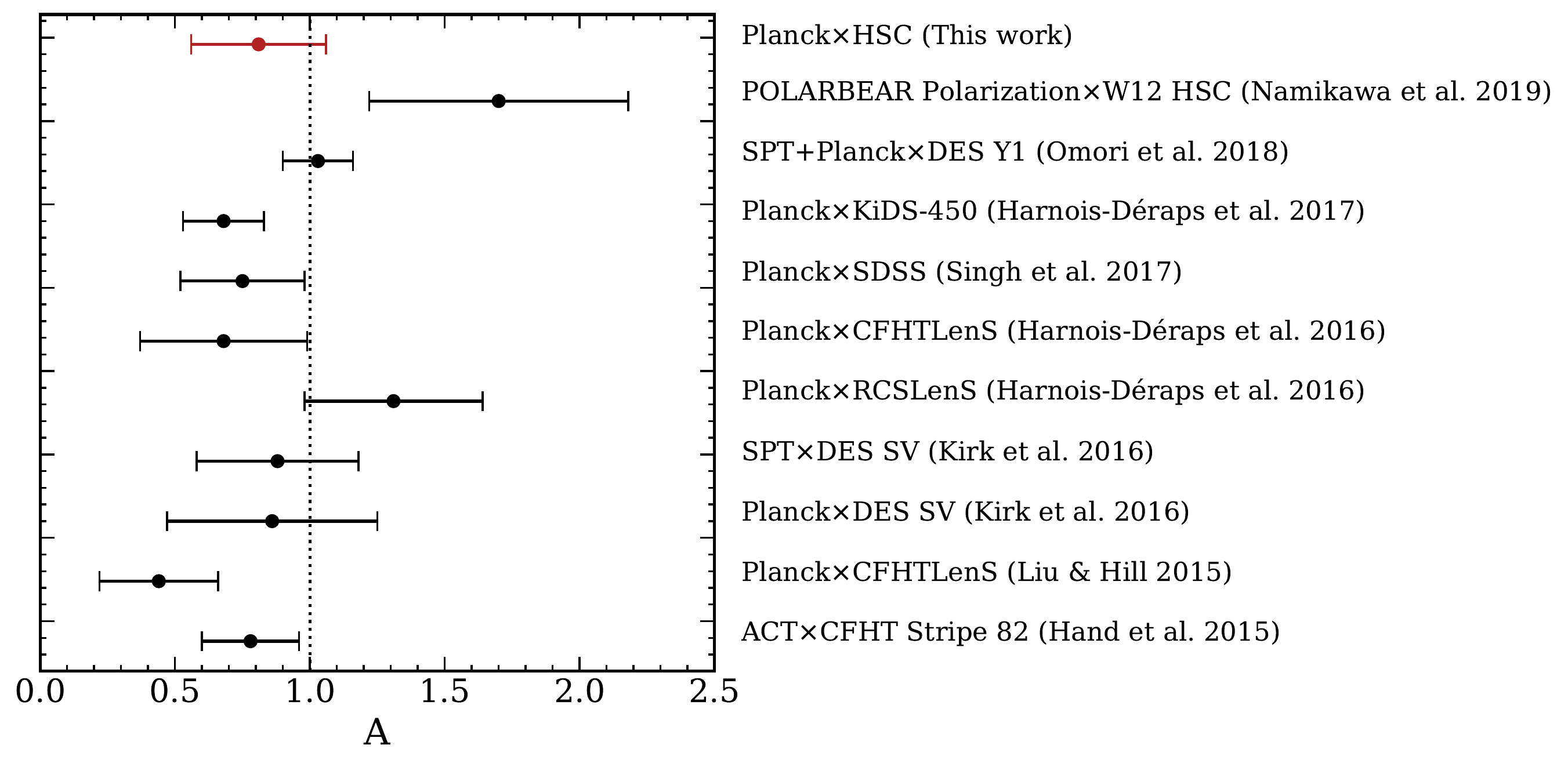}
	\caption{Best-fit amplitudes $A$ with $1\sigma$ confidence intervals from this work (red) and recent works studying cross-correlations between CMB lensing and galaxy weak lensing (black).}
\label{fig:As_literature}
\end{figure*}

Our best fit amplitude is consistent with Planck 2018 cosmology ($A=1$), and in general agreement with all previous CMB lensing and galaxy WL cross-correlation measurements. We compare our results with previous works in Fig.~\ref{fig:As_literature}, with corresponding 1$\sigma$ errors: between POLARBEAR and HSC~\texttt{WIDE12H} WL field~\citep{namikawa2019evidence}, SPT+Planck and DES Y1~\citep{omori2019dark}, Planck and KiDS-450~\citep{harnois2017kids}, Planck and CFHTLenS~\citep{liu2015cross,harnois2016cfhtlens}, Planck and RCSLenS~\citep{harnois2016cfhtlens}, Planck and SDSS ~\citep{singh2017cross}, SPT and DES-SV~\citep{kirk2016cross} and  ACT and CS82~\citep{hand2015first}. 

Previously, \cite{namikawa2019evidence} studied the cross correlation between one of the HSC fields, \texttt{WIDE12H}, and CMB lensing from the polarization data of POLARBEAR. Their best fit $A$ (1.70$\pm$0.48) is higher than our finding by $\sim1.2\sigma$ (statistical errors only). For a direct comparison, we estimate the amplitude for $\texttt{WIDE12H}$ field only and found $A^{\rm{W12}} = 1.05\pm 0.67$ and SNR=1.5, with $\chi^{2,\rm{W12}}_{\rm min}$=4.3 for DOF=4~(PTE= 36.6\%). Our test result is consistent within $1\sigma$ sigma with \cite{namikawa2019evidence} as well as Planck 2018. The large errors prevent a more precise comparison at present. However, the constraints will be significantly improved with future data from HSC, which is expected to cover 1400~deg$^2$ of sky in total~\citep{aihara2018hyper}.

\section{Systematic and Null Tests}
\label{sec:null_tests}
\begin{figure*}
	\centering
	\includegraphics[scale=0.4]{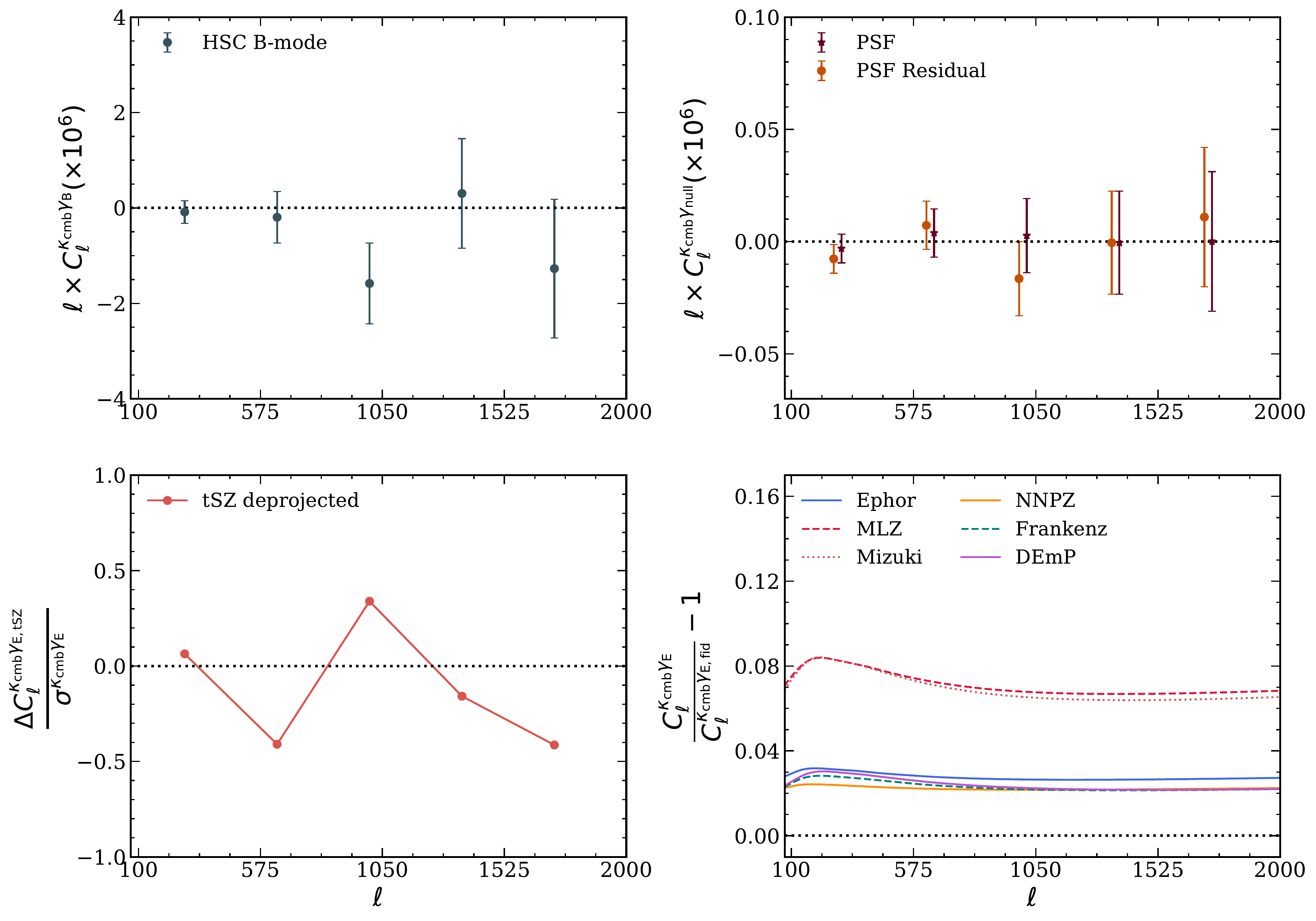}
	\caption{Systematic and null tests for the cross-power spectra, shown as an inverse-variance weighted average of the six HSC fields. {\bf Upper-left panel}: cross-correlation of Planck lensing with HSC B-mode maps. {\bf Upper-right panel}: cross-correlation of Planck lensing maps with HSC PSF leakage (red) and PSF residual (orange) maps. 
	{\bf Lower-left panel}: difference between cross-correlation using the default MV Planck lensing map and that with the tSZ-deprojected Planck lensing map, in units of the statistical error.  
	{\bf Lower-right panel}: fractional differences between the theoretical cross-power spectra assuming different photo-$z$ codes from that in our baseline analysis (\texttt{Ephor AB}).} 
\label{fig:cl_nulls}
\end{figure*}

We perform several consistency checks to test for systematic residuals and to verify our results. Our results are summarized in Table~\ref{tab:null_chisq} and Fig.~\ref{fig:cl_nulls}.

\begin{table}
\begin{tabular}{lcc}
\hline
Null Test     & $\chi^2_{\rm null}$/DOF & PTE ($\%$)  \\ \hline
B-mode   & 0.7               & 77.2 \\
Rotation & 0.7              & 82.3 \\
PSF leakage     &  0.9           & 56.6 \\
PSF residual & 1.0                & 47.1 \\ \hline
\end{tabular}
\caption{\label{tab:null_chisq}The $\chi_{\rm null}^2$ and PTE values for null tests of B-mode, Rotation of the ellipticities, PSF leakage, and PSF residuals.}
\end{table}

\subsection{HSC Lensing B-mode}

Because gravitational lensing can only generate E-modes (to lowest order), a non-vanishing B-mode signal is a diagnostic for systematics in the measurements. B-mode maps can be generated by taking the imaginary part of the \citet{kaiser1993mapping} inversion. We then follow the same masking and smoothing procedure to obtain the B-mode cross-correlations. We cross-correlate the B-mode WL map with 300 simulated Planck CMB lensing maps to obtain the B-mode covariance matrix.

We show the B-mode signal in the upper-left panel of Fig.~\ref{fig:cl_nulls}. We find the best-fit amplitute to be $A_{\rm B} = -0.24\pm 0.33$, consistent with 0. A similar test can be done by randomly rotating the galaxies in the HSC catalog. This effectively removes the lensing signal in galaxy WL maps. We repeat this procedure 300 times, i.e., 300 realizations of random rotations. The signal from the average of the 300 cross-power spectra is also consistent with zero. These test results are summarized in Table~\ref{tab:null_chisq}.

\subsection{HSC PSF}
We test for PSF leakage and PSF residuals following \citet{bosch2018hyper} and \citet{osato2020cross}. For PSF leakage, we build the maps using the estimated PSF~($e_{\rm p}$), instead of galaxy ellipticities, at the position of source galaxies. For PSF residuals, we build the maps using the differences of estimated PSFs and  the PSFs  measured from stars (the ``true'' PSF), $e_{\rm residual} = e_{\rm p}-e_{\rm star}$. To construct these maps, we adopt equal weights since all stars and the PSFs have similar signal-to-noise. To estimate the errors, we randomly rotate the PSFs  and PSF residuals and generate 300 realizations each.

The results are shown in the upper-right panel of Fig.~\ref{fig:cl_nulls} and summarized in Table~\ref{tab:null_chisq}. The signals from both tests are consistent with zero, validating that there is negligible impact from either the PSF or the PSF residuals.

\subsection{Planck tSZ}

The tSZ effect in the CMB is due to inverse-Compton scattering of CMB photons off free electrons that are primarily located in the hot intracluster medium  of galaxy clusters.  If the tSZ signal is not properly removed, it can bias CMB lensing reconstruction due to its non-Gaussianity and its correlation with large-scale structure ~\citep{van2014cmb,Osborne:2013nna,novaes2016local,hojjati2017cross,madhavacheril2018mitigating,schaan2019foreground}. We test for effects due to tSZ contamination in the Planck MV lensing potential maps using the  \texttt{COM\_Lensing-Szdeproj\_4096\_R3.00} map from Planck, reconstructed from the tSZ-deprojected \texttt{SMICA} temperature map~\citep{lensing2018planck}.

The deviation of the cross-power spectrum obtained using the tSZ-deprojected Planck lensing map from that using the Planck MV lensing map (as done in our main analysis) is shown in the lower-left panel of Fig.~\ref{fig:cls_cross_result}, in units of the standard deviation of the latter. The two measurements are consistent within 0.5$\sigma$ in all multipole bins, confirming that tSZ contamination to our measurements is negligible.

\subsection{HSC Photo-z}
\label{sec:photo-z}
A biased photo-$z$ distribution can impact the theoretically predicted cross-power spectrum. 
HSC adopts seven photo-$z$ algorithms: \texttt{MLZ, Ephor, Ephor AB, Mizuki, NNPZ, Frankenz} and \texttt{DEmP}. The estimated $p_{\rm z}$ from each code is shown in Fig.~\ref{fig:pz_stacked}. We estimate the impact of photo-$z$ error by comparing the theoretical model predicted by each photo-$z$ code to that used in our baseline analysis (\texttt{Ephor AB}). 

We show our results in the lower-right panel of Fig.~\ref{fig:cl_nulls}. Most algorithms produce small differences with respect to our baseline model ($\lesssim 3\%$), with an exception of the \texttt{MLZ} and \texttt{Mizuki} code (up to $\sim$ 8\%). Taking these two cases, we repeat our analysis and we obtain $A =0.86 \pm 0.28$ for \texttt{MLZ} (SNR=3.1, $\chi^2_{\rm min} = 31.3$, PTE= 34.4\%)  and $A =0.87 \pm 0.27$ for \texttt{Mizuki} (SNR=3.1, $\chi^2_{\rm min} = 31.3$, PTE= 34.9\%). Both are consistent with our main analysis results. Therefore, we conclude that the choice of photo-$z$ algorithm does not impact significantly our conclusion. 

We further test the impact of photometric outliers---galaxies falsely assigned much higher or lower redshifts than their actual redshifts. \textred{
Outliers are expected to occur more frequently at high redshift where it is harder to obtain features of the UV continuum and to deal with Lyman break degeneracy.} This will inevitably impact the cross-correlations as the WL and CMB lensing kernels overlap the most in the high-redshift tails of $p_z$ (as shown in Fig.\ref{fig:pz_stacked}). To test the potential impact of photo-$z$ outliers, we manually reassign a fraction of galaxies as outliers and modify the $p_z\to (1-f_{\rm out})p_z+f_{\rm out}p_{\rm out}$, where $f_{\rm out}$ is the outlier fraction and $p_{\rm out}$ is taken to be the stacked photo-$z$ PDF of all \textred{$z_{\rm best}>1.5$} galaxies. \citet{tanaka2018photometric} estimated approximately $15\%$ outliers for HSC galaxies at $i_{\rm lim}=25$ and $8\%$ at $i_{\rm lim}=24$. With our magnitude cut of $i_{\rm lim}=24.5$, we consider 3 scenarios, $f_{\rm out}$=15\% (pessimistic), 8\% (realistic), and 5\% (optimistic). 
We compute the theoretical model for each case and find that the best-fit amplitude $A$ decreases by 21\%, 12\%, and 7\% for the pessimistic, realistic, and optimistic cases, respectively. In all cases, the bias due to photo-$z$ outliers is sub-dominant compared to our statistical uncertainty and therefore is unlikely to significantly affect our results. \textred{We must also mention that while the photo-z uncertainties are sub-dominant for our analysis, this will change with the upcoming surveys and therefore must be well understood to achieve the scientific goals of a Stage-IV galaxy WL surveys.}

\subsection{HSC Multiplicative Bias}

Similar to the photo-$z$ measurements, the precision of multiplicative biases in shape measurements also decreases at higher redshift. \citet{mandelbaum2018first} tested extensively and controlled the multiplicative bias at the 1\% level. However, any small fluctuation at high redshift might become significant in the cross-correlation signal. Therefore, the galaxy WL-CMB lensing cross-correlation allows us to calibrate these systematics as the parameter $A$ also absorbs this information \citep{vallinotto2012using,das2013cmb,Liu2016multiplicative}. Given the precision of our current measurements, we expect this type of calibration to be noisy compared to the amplitude of these systematics. Nevertheless, it is important to perform this consistency check as it will become crucial with  upcoming data. In addition, this is a complementary test of the photo-$z$ bias discussed in Sec.~\ref{sec:photo-z}.

\begin{figure}[h!]
	\centering
	\includegraphics[width=\linewidth]{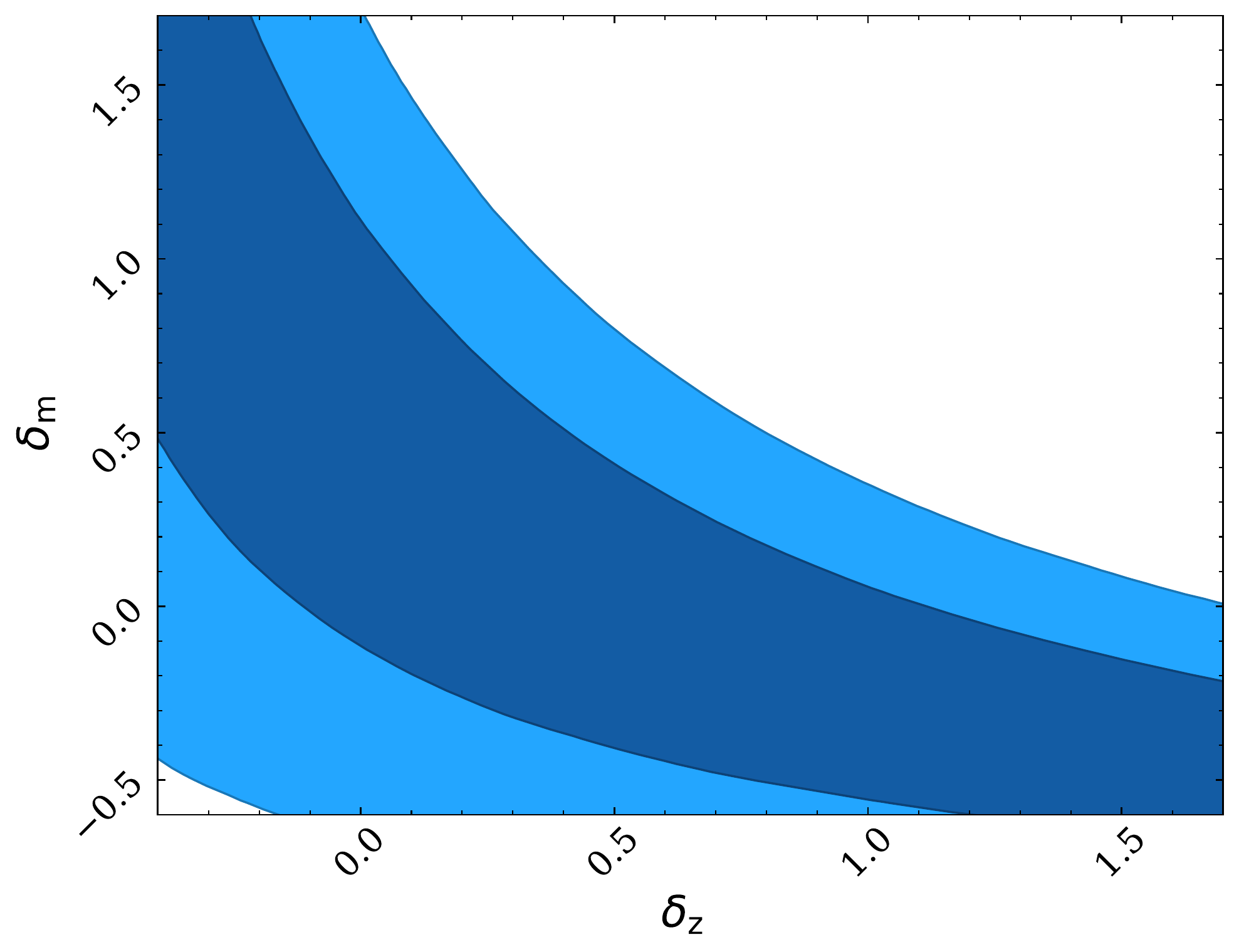}
	\caption{Shear multiplicative bias correction $\delta_{\rm m}$ and photometric redshift distribution correction $\delta_{\rm z}$ constraints. The contour regions represents 68\% and 95\% CLs.}
\label{fig:m_z_calib}
\end{figure}

To calibrate the joint multiplicative and photo-$z$ biases at high redshift, we repeat the same analysis but on galaxies with $z_{\rm best}>1.5$. The signal for this high-$z$ sample is measured at SNR=5.8 and the best-fit amplitude is $A_{\rm high} = 1.21 \pm 0.20$, with PTE= $34.8\%$. We also carry out a B-mode null test, finding the signal consistent with zero, with $\chi
^2_{\rm null}/{\rm DOF} = 1.1$ and PTE = $30.8$\%.
Following \citet{harnois2017kids}, we model the shift in the amplitude due to the additional multiplicative shear bias $\delta_{\rm m}$ and the photo-$z$ $\delta_{\rm z}$  as $(1+\delta_{\rm m})(1+\delta_{\rm z}) = A_{\rm high}/A_{\rm main}$, where $A_{\rm main}$ is the amplitude measured from our main analysis with redshift cuts $0.3<z_{\rm best}<1.5$. 
We assume wide flat priors for the two parameters. We show the constraints on $\delta_{\rm m}$ and $\delta_{\rm z}$ in Fig. \ref{fig:m_z_calib}. $\delta_{\rm m}$ and $\delta_{\rm z}$ are consistent with 0 within 95\% CL.

While our current measurements do not put stringent constraints on the multiplicative and photo-$z$ biases, \citet{schaan2017} forecasted such analysis for future surveys and found a joint analysis of Stage-IV galaxy WL surveys and CMB-S4 can calibrate the shear bias at sub-percent level and hence is a complementary tool to the WL image simulations that have been typically done to date.

\subsection{Intrinsic Alignments}
 
The intrinsic alignment (IA) due to tidal interactions of foreground galaxies also correlates with the CMB lensing field, generating a secondary signal in our measured correlation, 
\begin{align}
    C_{\ell}^{\rm obs,\kappa_{\rm cmb}\gamma_{\rm E}} = C_{\ell}^{\kappa_{\rm cmb}\gamma_{\rm E}} +C_{\ell}^{\kappa_{\rm cmb}\rm IA}.
\end{align}
Here, the additional term $C_{\ell}^{\kappa_{\rm cmb}\rm IA}$ represents the cross-correlation between the IA of galaxies and the CMB lensing field.\ 

The IA contribution has been estimated to reduce the cross-correlation amplitude by 5--20$\%$, depending on the redshift distribution and the type of galaxy~\citep{joachimi2015galaxy,chisari2015contamination,troxel2015intrinsic,kirk2016cross,larsen2016intrinsic}. This is because galaxies tend to align radially from the gravitational potential, opposite of the lensing signal. 

We adopt a nonlinear alignment model (NLA)  \citep{hirata2004intrinsic,bridle2007dark} to examine the potential contamination by IA in our results. In this model, the $C_{\ell}^{\kappa_{\rm cmb}\rm IA}$ term can be calculated by replacing the galaxy lensing kernel in Eq.~\ref{eq:kernel_gal} with
\begin{align}
    W^{\rm IA}(\chi) = \frac{-C_1 \rho_{\rm crit} \Omega_{\rm m} }{D(z)} p_{\rm z}(z)\frac{d z}{d\chi},
\end{align}
where $\rho_{\rm crit}$ is the critical density at $z=0$ and $D(z)$ is the linear growth function, normalized to unity at $z=0$. The normalization factor $C_1$ is fixed at $5\times 10^{-14} h^{-2}\textup{M}_\odot^{-1}\rm Mpc^3$, based on the SuperCOSMOS  measurement at low redshift \citep{brown2002measurement}. We make simple assumptions of no redshift evolution and no galaxy-morphology dependence, as limited by the level of signal in our measurements. 

With the inclusion of the additional IA term, we obtain a best fit amplitude $A= 0.89\pm 0.27$ (compared with $A= 0.81\pm 0.25$ in our main analysis) and the significance of the detection remains unchanged at $3.1\sigma$. \textred{We conclude that, within our current observational constraints and galaxy alignment model, our results are unaffected.  We will need to include an accurate IA model in analysis of future data.}

\subsection{Other Systematics}
There are other potential sources of contamination, such as the kinematic Sunyaev–Zel'dovich~(kSZ) effect and baryonic processes. The kSZ effect originates from the Doppler shift of CMB photons induced by Compton-scattering off electrons in bulk motion.  Because it preserves the blackbody spectrum of the CMB, it cannot be removed by the multifrequency foreground separation technique as typically done to separate the tSZ signal. \citet{FerraroHill2018} forecasted that the kSZ bias in a cross-correlation of Planck CMB lensing and Rubin Obs. galaxy WL would be $\approx$5\% for CMB lensing reconstruction using modes up to $\ell_{\rm max}$=3000. At the noise levels of the data used in our analysis, the contamination is at the  $\lesssim 1\%$ level and is negligible compared to our errors. 

Baryonic processes such as black hole accretion, radiative cooling, star formation, and supernova feedback redistribute the gas inside a halo and hence reshape the gravitational potential. From a combination of observations and hydrodynamic simulations, we expect that the matter power is suppressed by $\approx$10\% at scales $k\gtrsim 1\; h {\rm Mpc}^{-1}$~\citep{hellwing2016effect,springel2018first,chisari2018impact}. Consequently, baryonic processes can impact the galaxy lensing-CMB lensing cross-correlation, causing a decrease of the measured signal.  However, for the main redshift range we are probing ($z\approx1$), $k\gtrsim 1 h {\rm Mpc}^{-1}$ corresponds to $\ell\gtrsim 5000$, which is already excluded by our scale cut of $\ell_{\rm max}=1900$. It is worth mentioning that to estimate the baryonic impact, \citet{harnois2016cfhtlens, harnois2017kids} considered a model built from the OWLS  hydrodynamical simulations~\citep{schaye2010physics,van2011effects} and found the suppression in the matter power spectrum to be 10--20\% at $\ell=$1000-2000. \textred{Even with this scenario, the level of suppression in our analysis due to baryons remains well within our errors, but such effects must be included in more accurate future measurements.}

\section{Conclusions} 
\label{sec:conclusions}
In this work, we study the cross-correlations between Planck CMB lensing and HSC galaxy weak lensing maps, with total unmasked overlapping sky coverage of  $\approx 131\deg^2$. With source galaxies between redshift 0.3 and 1.5, this measurement allows us to probe gravitational potentials at redshift $z\approx 0.8$. 

The cross-correlation signal is measured at $3.1\sigma$. We model the signal  as an amplitude $A$ with respect to the Planck 2018 cosmology, obtaining $A= 0.81\pm 0.25$. Our results are fully consistent with Planck cosmology ($A=1$) as well as previous cross-correlation analyses using Planck CMB lensing maps and other galaxy weak lensing surveys~\citep{hand2015first,liu2015cross,kirk2016cross, harnois2016cfhtlens,singh2017cross,harnois2017kids, omori2019dark,  namikawa2019evidence}. 

We perform various systematic tests, including weak lensing B-modes, rotated galaxy ellipticities, PSF leakage, PSF residual, photometric redshift errors and outliers, tSZ leakage, multiplicative shear bias, and intrinsic alignments. We find no significant systematic residuals and our measurement is robust to these effects.

The CMB lensing and galaxy weak lensing cross-correlation offers an excellent opportunity to constrain cosmology, test gravity, and calibrate systematics across a wide range of redshifts. It is a complementary probe to the auto-correlation analyses. Together with previous works, we demonstrate the feasibility and robustness of this technique. Current studies are somewhat limited by the large noise in CMB lensing and small sky coverage by the galaxy weak lensing survey. We expect significant improvements from combining ongoing and future CMB surveys with lower noise such as Advanced ACT~\citep{Henderson:2016fxx}, the Simons Observatory~\citep{ade2019simons}, SPT-3G~\citep{benson2014spt}, and CMB-S4~\citep{abazajian2016cmb}, and large scale galaxy weak lensing surveys such as the Vera Rubin Observatory LSST\footnote{\url{https://lsst.slac.stanford.edu/}}~\citep{abell2009lsst}, Euclid\footnote{\url{https://sci.esa.int/web/euclid/}}~\citep{laureijs2011euclid}, and the Nancy Grace Roman Space Telescope\footnote{\url{https://roman.gsfc.nasa.gov/}}, formerly known as the Wide Field Infrared Survey Telescope (WFIRST) ~\citep{green2012wide}.

\acknowledgments

GM and KMH acknowledge support from National Science Foundation award 1815887. This work is supported by an NSF Astronomy and Astrophysics Postdoctoral Fellowship (to JL) under award AST-1602663. JCH thanks the Simons Foundation for support.
We acknowledge the use of many public python packages: Numpy \citep{oliphant2015guide}, healpy \citep{zonca2019healpy}\footnote{\url{https://healpy.readthedocs.io/en/latest/index.html}}, Astropy\footnote{\url{http://www.astropy.org}} a community-developed core Python package for Astronomy~\citep{astropy:2013, astropy:2018}, Matplotlib~\citep{hunter2007matplotlib}, emcee~\citep{foreman2013emcee}, flipper \citep{das2009efficient}, IPython~\citep{perez2007ipython} and Scipy~\citep{jones2001scipy}.

\vspace{5mm}

\bibliography{hsc_cross}
\bibliographystyle{aasjournal}

\end{document}